\documentclass{pasj01}
\Received{$\langle$reception date$\rangle$}
\Accepted{$\langle$acception date$\rangle$}
\Published{$\langle$publication date$\rangle$}

\begin{document}
\title{The Origin of Recombining Plasma and the Detection of the Fe-K Line in the Supernova Remnant W28}
\author{Hiromichi \textsc{Okon}\altaffilmark{1}}
\author{Hiroyuki \textsc{Uchida}\altaffilmark{1}}
\author{Takaaki \textsc{Tanaka}\altaffilmark{1}}
\author{Hideaki \textsc{Matsumura}\altaffilmark{1}}
\author{Takeshi~Go \textsc{Tsuru}\altaffilmark{1}}%
\altaffiltext{1}{Department of Physics, Faculty of Science, Kyoto University, Kitashirakawa Oiwake-cho, Sakyo-ku, Kyoto, Kyoto 606-8502, Japan}
\email{okon@cr.scphys.kyoto-u.ac.jp}

\KeyWords{ISM: individual objects (W28, G6.4$-$0.1) --- ISM: supernova remnants --- cosmic rays --- X-rays: ISM}

\maketitle

\begin{abstract}
Overionized recombining plasmas (RPs) have been discovered from a dozen of mixed-morphology (MM) supernova remnants (SNRs).
However their formation process is still under debate. 
As pointed out by many previous studies, spatial variations of plasma temperature and ionization state provide clues to 
understand the physical origin of RPs.
We report on a spatially resolved X-ray spectroscopy of W28, which is one of the largest MM SNRs found in our Galaxy.
Two observations with Suzaku XIS cover the center of W28 to the northeastern rim where the shock is interacting with molecular clouds.
The X-ray spectra in the inner regions are well reproduced by a combination of two-RP model with different temperatures and ionization states, whereas that in northeastern rim is explained with a single-RP model.
Our discovery of the RP in the northeastern rim suggests an effect of thermal conduction between the cloud and hot plasma, which may be the production process of the RP.
The X-ray spectrum of the northeastern rim also shows an excess emission of the Fe\emissiontype{I} K$\alpha$ line.
The most probable process to explain the line would be inner shell ionization of Fe in the molecular cloud by cosmic-ray particles accelerated in W28.
\end{abstract}

\section{Introduction}
W28 (a.k.a., G6.4$-$0.1) is one of the most well-studied Galactic supernova remnants (SNRs) interacting with molecular clouds \citep{Wootten1981}.
The remnant is also known as one of the typical mixed-morphology (MM) SNRs which are characterized by a radio shell with center-filled X-rays \citep{Rho1998}.
Due to its  old age (33--42~kyr; \cite{Kaspi1993,Velazquez2002,Li2010}) and a relatively small distance ($\sim2$~kpc based on the H\emissiontype{I} observation; \cite{Velazquez2002}), W28 has a large apparent size of $\sim48\arcmin$ \citep{Seward1990}, which enables us to investigate the small physical scale structures associated with this SNR.

W28 appears to be expanding into a dense interstellar medium (ISM) as suggested by a global structure of H\emissiontype{I} gas \citep{Velazquez2002}.
\citet{Nobukawa2018} found the Fe\emissiontype{I} K$\alpha$ emission roughly coincident  with the H\emissiontype{I} distribution, indicating possible interactions between low-energy cosmic-ray protons (LECRp) and the dense ISM in the vicinity of W28.
Also, as first suggested by \citet{Wootten1981},  many previous studies provided evidence that the northeastern part of W28 is interacting with a dense molecular cloud.
These studies include detections of OH maser (1720~MHz) spots \citep{Frail1994,Claussen1997}, the presence of shocked gas traced by $\rm{^{12}CO\,({\it{J}}=3-2)}$ and $\rm{^{12}CO\,({\it{J}}=1-0)}$ lines \citep{Arikawa1999}, and GeV/TeV gamma-ray emission coincident with these  molecular clouds \citep{Aharonian2008,Abdo2010,Giuliani2010}.

Past observations of W28 with ROSAT and ASCA revealed a centrally concentrated X-ray emission as well as  a bright northeastern X-ray rim \citep{Rho2002}.
Recently, \citet{Sawada2012} observed the central region of W28 with Suzaku and discovered an overionized recombining plasma (RP), which was supported by a Chandra observation of the same region \citep{Pannuti2017}.
While the formation process of the RP is poorly understood, one of the possible scenarios is a thermal conduction between SNRs and surrounding molecular clouds \citep{Kawasaki2002}.
Although \citet{Sawada2012} prefer another possible scenario ``rarefaction'' \citep{Itoh1989,Shimizu2012},  a more comprehensive spatially resolved study is required to approach the true origin of the RP in W28.

The gas environment in the vicinity of SNRs is a key to understanding the physical origin of the RPs for both scenarios \citep{Itoh1989,Kawasaki2002}.
Supportive evidence for the thermal conduction scenario is found from two RP SNRs, G166.0$+$4.3 \citep{Matsumura2017a} and IC~443 \citep{Matsumura2017b}.
Their main argument is that X-ray emitting plasmas surrounded by dense ISM or molecular clouds are significantly cooled and are in the recombining state.
In this context, the bright northeastern rim of W28 is also expected to be overionized.
\citet{Nakamura2014} and \citet{Zhou2014}, however, reported no detection of an RP from the X-ray spectrum of the northeastern rim obtained with XMM-Newton, which may conflict with the thermal conduction scenario.
Since the RP has been robustly detected from the central region by \citet{Sawada2012}, it is required to investigate the spatial variation of temperature and ionization state from the center toward the northeastern rim.

In this paper, we report on the result of a spatially resolved analysis of the central to the northeastern parts of W28 with Suzaku.
The Suzaku observation of the northeastern part is analyzed in our paper for the first time.
Strong K-shell emissions from neutral and highly ionized Fe are detected from the northeastern bright rim and the center of W28, respectively.
Both are key to understanding the formation process of RPs and the interaction between the northeastern rim and surrounding molecular clouds.
Throughout the paper, errors are quoted at 90\% confidence levels in the tables and text.
Error bars shown in the figures are 1$\sigma$ confidence levels.
The distance to W28 is assumed to be 2~kpc \citep{Velazquez2002}.

\section{Observations and data reduction}
Two-pointing observations (the center and the northeast) of W28 were carried out with Suzaku in 2010 April and 2011 February, respectively.
The observation log is summarized in table~\ref{tab:obs_log}.
We analyzed data taken with the X-ray Imaging Spectrometer (XIS; \cite{Koyama2007}) aboard Suzaku.
The XIS is composed of four CCD cameras (XIS0, 1, 2, and 3), which are installed on the focal planes of the X-Ray Telescopes (XRTs; \cite{Serlemitsos2007}). 
XIS2 and a segment of XIS0 have malfunctioned and been out of operation since 2006 November and 2009 June, respectively (Suzaku XIS documents\footnote{ftp://legacy.gsfc.nasa.gov/suzaku/doc/xis/suzakumemo-2007-08.pdf}${^,}$\footnote{ftp://legacy.gsfc.nasa.gov/suzaku/doc/xis/suzakumemo-2010-01.pdf}).

We reduced the data with the HEADAS software version 6.19 and the calibration database released in 2016 June.
We selected events with the standard event selection criteria provided by the Suzaku team.
We estimated the non-X-ray background (NXB) with $\texttt{xisnxbgen}$ (\cite{Tawa2008}).
We generated the redistribution matrix files and the ancillary response files by using $\texttt{xisrmfgen}$ and  $\texttt{xissimarfgen}$ (\cite{Ishisaki2007}), respectively.
We used version 12.9.0u of the XSPEC software \citep{Arnaud1996} for the following spectral analysis.
For analyzing the XIS0 and XIS3 data, we ignored the energy band of 1.7--2.0~keV because of the calibration uncertainty around the neutral Si K-edge (Suzaku XIS documents\footnote{https://heasarc.gsfc.nasa.gov/docs/suzaku/analysis/sical.html}).

\begin{table*}
 \tbl{Observation log.}{
 \begin{tabular}{llllll}
     \hline
     Target & Obs. ID & Obs. date & $\rm{(R.A. , Dec.)}^{\ast}$ & Effective exposure~(ks) \\      
     \hline
     	W28\_CENTER & 505005010 & 2010-Apr-03 & (\timeform{18h00m17s},~\timeform{-23D21'59''}) & 73.0 \\
	W28\_NE & 505006010 & 2011-Feb-25 & (\timeform{18h01m30s},~\timeform{-23D17'30''}) & 100 \\
	      \hline
   \end{tabular}}
   \label{tab:obs_log}
\begin{tabnote}
${\ast}$ Equinox in J2000.0.\\
\end{tabnote}
\end{table*}

\section{Analysis and results}

\subsection{Image}

\begin{figure*}
\begin{center}
\includegraphics[width=14cm]{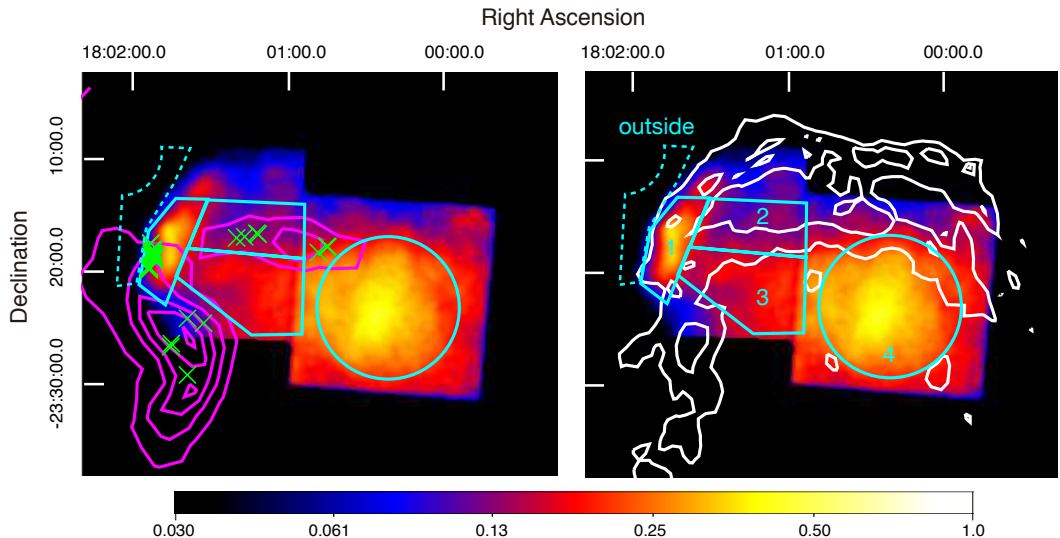} 
\end{center}
\caption{Images of W28 in the energy band of 0.65--4.0~keV after NXB subtraction and vignetting effect and exposure correction.
The cyan solid and dash lines denote the source and background regions, respectively.
In the left panel, the magenta contours indicate the distribution of the $\rm ^{12}CO~({\it J}=2-1)$ line emissions for $V_{\rm LSR}$ = --15 km s$^{-1}$ to 15 km s$^{-1}$ as observed with the NANTEN2 telescope \citep{Torii2011}.
The magenta contours are drawn every 40 K km s$^{-1}$ from 40 K km s$^{-1}$.
The 1720~MHz OH maser emission \citep{Claussen1997} detected at the locations indicated by the green crosses are taken with the Karl G. Jansky Very Large Array (VLA).
In the right panel, the radio 325~MHz image shown as the white contours are taken with VLA.
The white contours are drawn  every 0.225 Jy/beam from 0.05 Jy/beam.
}
\label{fig:W28_Xray_molecularclouds}
\end{figure*}

Figure~\ref{fig:W28_Xray_molecularclouds} shows an NXB-subtracted and vignetting and exposure corrected image of W28 (0.65--4.0~keV).
The centrally peaked morphology is seen at the SNR's center, where \citet{Sawada2012} found the RP.
The rim-brightened partial shell \citep{Rho2002} is present in the northeastern region, where \citet{Nakamura2014} and \citet{Zhou2014} found no evidence for an RP.
The interior region between the center and the northeastern rim can further be divided into two parts: the northern part where a shock-cloud interaction is suggested from the $^{12}$CO and OH maser emissions and the southern part where molecular clouds were not found through past observations.
In order to investigate the spatial association between plasma parameters and the ambient molecular clouds, we thus selected four regions which are enclosed by the cyan lines in figure~\ref{fig:W28_Xray_molecularclouds}.

\subsection{Fe lines}

\begin{figure}
\begin{center}
\includegraphics[width=7.5cm]{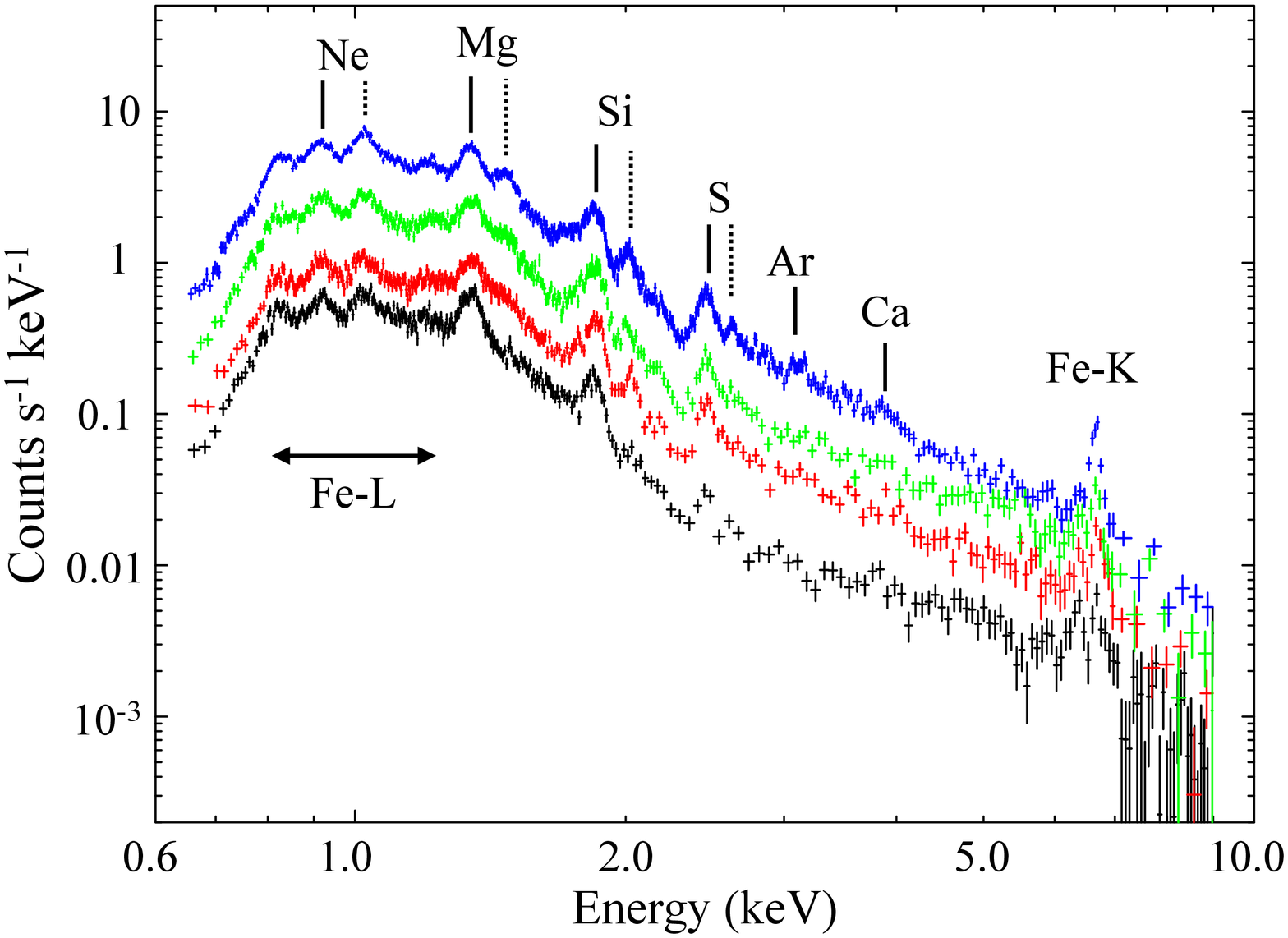} 
\includegraphics[width=7.5cm]{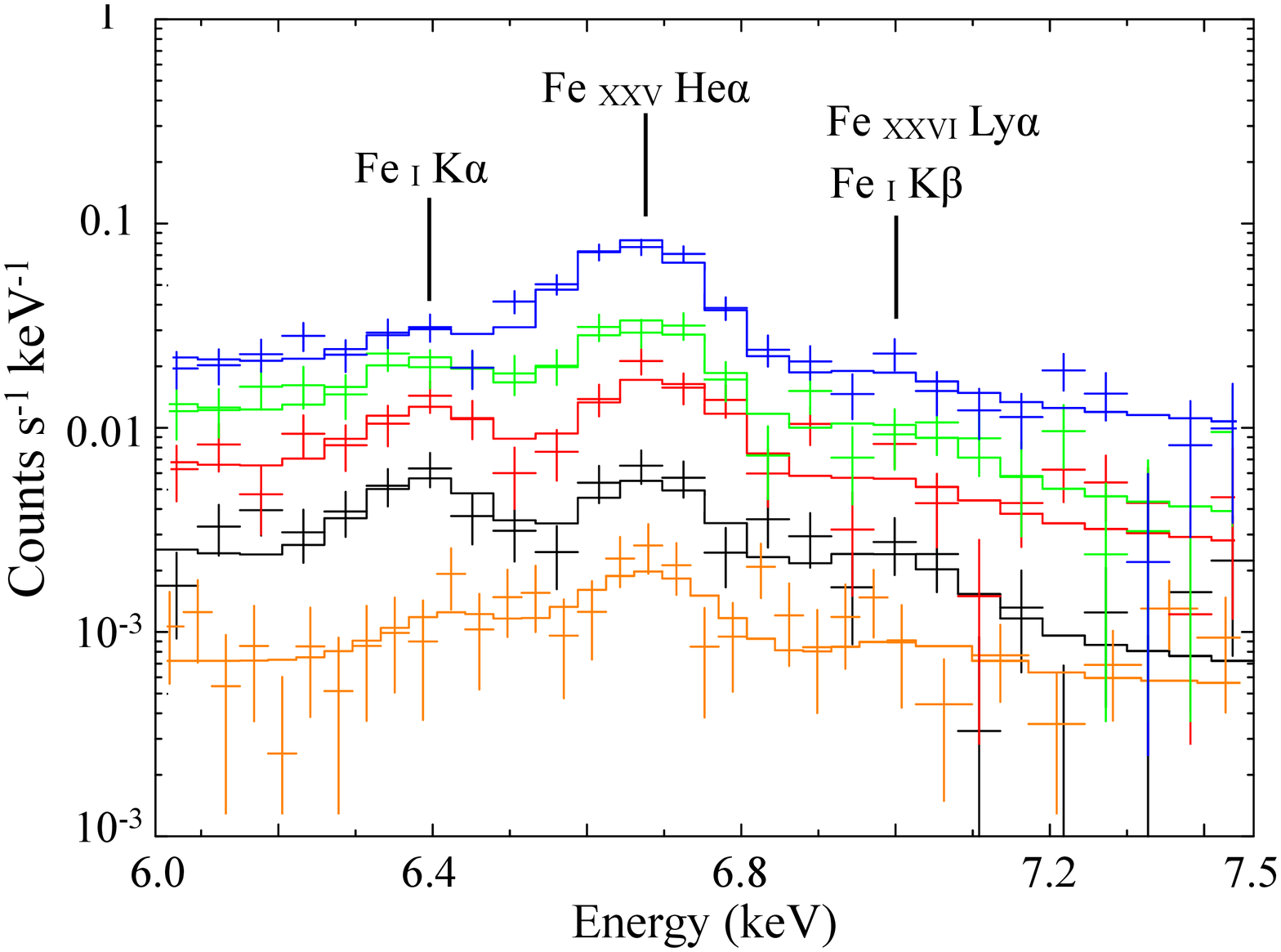} 
\end{center}
\caption{Spectra of regions outside (orange), 1 (black), 2 (red) , 3 (green), and 4 (blue) in the energy band of 0.65--4.0~keV (upper) and 6.0--7.5~keV (lower) obtained with the XIS0 $+$ 3.
NXB was subtracted. 
The spectra of regions outside, 2, 3, and 4 are scaled by factors of 0.5, 2.0, 3.0, and 2.0, respectively, for display purpose. 
The vertical solid and dashed black lines in the top  panel represent the centroid energies of the He$\alpha$ lines, Ly$\alpha$ lines, respectively.
The vertical solid lines in the bottom panel denote the centroid energies of the Fe-K ines.}
\label{fig:3region_spectra}
\end{figure}

W28 is located near the Galactic center, where the Galactic ridge X-ray emission (GRXE) is not negligible especially in the high energy band.
Figure~\ref{fig:3region_spectra} shows NXB-subtracted XIS spectra taken from the four regions.
We detected emission lines of Fe K$\alpha$ in the 6.0--7.0~keV band as well as the Ne, Mg, Si and S K $\alpha$ lines in the low energy band.
As shown in the lower panel of figure~\ref{fig:3region_spectra}, there are two prominent lines around $\sim6.5$~keV.
The measured line centroids are 6370$^{+30}_{-15}$~eV and 6671$^{+11}_{-10}$~eV, which are consistent with those of Fe\emissiontype{I} K$\alpha$ (6.40~keV) and Fe\emissiontype{XXV} He$\alpha$ (6.68~keV), respectively.
It is reasonable to consider that they are \textit{mostly} attributed to the GRXE since the SNR emission \citep{Sawada2012} and a possible non-thermal emission including the Fe\emissiontype{I} K$\alpha$ line produced by LECRp  \citep{Nobukawa2018} are both relatively weak as noted by these studies.
We, however, carefully evaluated the contributions of each component as described below. 

\begin{figure}
\begin{center}
\includegraphics[width=8cm]{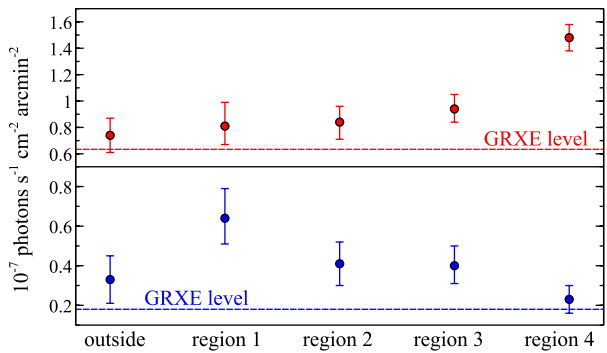} 
\end{center}
\caption{Intensities of Fe-K$\alpha$ lines of the five regions obtained by fitting the spectra of the five regions with the phenomenological model.
The red and blue points represent the intensities of Fe\emissiontype{XXV} He$\alpha$ line and the Fe \emissiontype{I} K$\alpha$ line, respectively.
The horizontal dashed lines indicate the GRXE level \citep{Yamauchi2016}.}
\label{fig:Fe_emission}
\end{figure}

Figure~\ref{fig:Fe_emission} presents the Fe\emissiontype{I} K$\alpha$ and Fe\emissiontype{XXV} He$\alpha$ intensities of each region measured by fitting the 6.0--7.5~keV spectra with a phenomenological model: a power-law continuum plus four Gaussians at 6.40~keV (Fe\emissiontype{I} K$\alpha$), 6.68~keV (Fe\emissiontype{XXV} He$\alpha$), 6.97~keV (Fe\emissiontype{XXVI} Ly$\alpha$), and 7.06~keV (Fe\emissiontype{I} K$\beta$).
We plot, for comparison, the intensities of a region outside W28 (``outside''; see figure~\ref{fig:W28_Xray_molecularclouds}).
We found that the intensities of both lines are almost consistent with the GRXE level in the outside region, whereas the Fe\emissiontype{I} K$\alpha$ and Fe\emissiontype{XXV} He$\alpha$ lines are relatively strong in region~1 and region~4, respectively.
As pointed out by \citet{Rho2002}, hard emission is concentrated on the center of the remnant.
We thus expect the presence of a high temperature plasma in region~4, which is able to emit the Fe\emissiontype{XXV} He$\alpha$ line.
On the other hand, the Fe\emissiontype{I} K$\alpha$ excess is significant only in region~1 where the shock is interacting with the clouds.
The result implies a similar Fe\emissiontype{I} K$\alpha$ enhancement reported for several Galactic SNRs \citep{Sato2014,Sato2016,Nobukawa2018}. 
This point will be discussed in more detail in section~\ref{sec:Fe_line}.

\subsection{Local X-ray Background}\label{sec:lxb}
We first estimated the background spectrum from a nearby blank-sky data (Obs ID: 500008010) following \citet{Sawada2012}.
However, the background level of the blank-sky data above 4~keV is significantly higher than the total flux of the W28 regions in the same energy band, although it is  within the global fluctuation of the GRXE presented by \citet{Uchiyama2013}.
Note that \citet{Sawada2012} only used the data below 5~keV for their spectral analysis.
We speculate that, in this case, the discrepancy of the GRXE levels between the two observations could be negligible.
On the other hand, we found significant contributions from the thermal emission from W28 around the Fe\emissiontype{XXV} He$\alpha$ line as explained above.
We therefore used another local background extracted from a source-free region of our observations, namely ``outside'' shown in figure \ref{fig:W28_Xray_molecularclouds}, in the following analysis.

\begin{figure}
\begin{center}
\includegraphics[width=8cm]{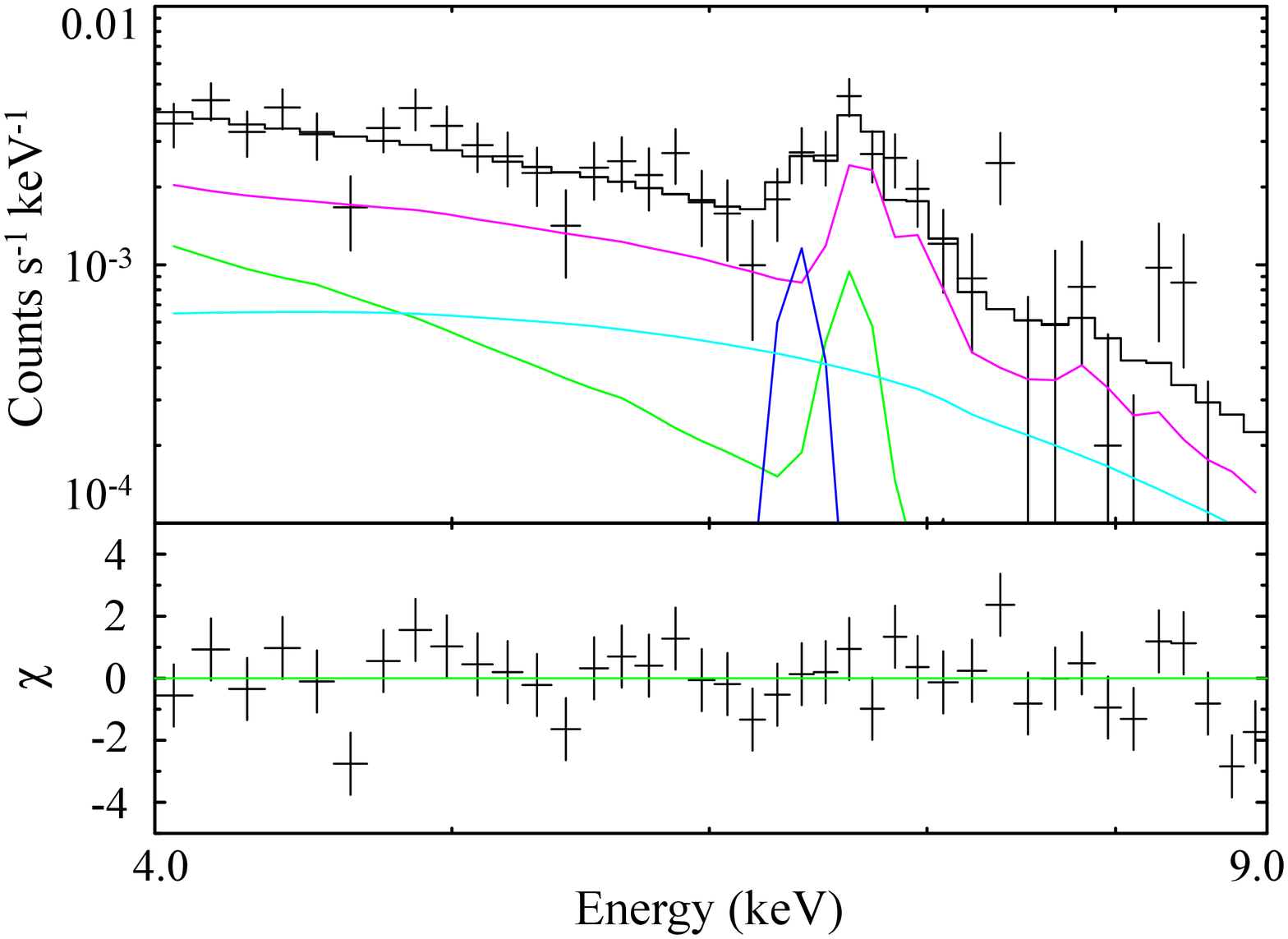} 
\end{center}
\caption{Spectrum of outside region in the energy band of 4.0--9.0~keV obtained with the XIS0 + 3.
The black lines represent the best-fit model, which are decomposed into each component indicated with the colored curves. 
The magenta, green, cyan and blue lines represent the HP, LP, CXB and Fe\emissiontype{I} K$\alpha$ line components, respectively.}
\label{fig:BG}
\end{figure}

Figure~\ref{fig:BG} shows the 4.0--9.0~keV spectrum obtained from the outside region.
We applied a background model consisting of the GRXE and the cosmic X-ray background (CXB).
The parameters for the CXB component were fixed to the values given by \citet{Kushino2002}.
Considering the results by \citet{Uchiyama2013} and \citet{Yamauchi2016}, we made a semi-phenomenological GRXE model composed of the foreground emission (FE), the high-temperature plasma  (HP) and the low-temperature plasma (LP) from the Galactic ridge and the Fe\emissiontype{I} K$\alpha$ line from the Galactic plane.
We fixed the parameters of the LP and HP to the values of \citet{Uchiyama2013}, except for $kT_e$ of the HP and normalization of the HP and LP.
The normalization of the HP was fixed at 0.29 times that of the LP \citep{Uchiyama2013}.
The electron temperature $kT_e$ and normalization of the LP were free parameters.
The intensity of the Fe\emissiontype{I} K$\alpha$ line was also allowed to vary, whereas that of the Fe\emissiontype{I} K$\beta$ to K$\alpha$ intensity ratio was fixed to 0.125 \citep{Kaastra1993}.
The best-fit model is shown in figure~\ref{fig:BG} and the  best-fit parameters are summarized in table~\ref{tab:bkg_model}.
We used this model as the background spectrum for W28.

\begin{table}
 \tbl{Best-fit model parameters of the background.}{
 \begin{tabular}{llll}
     \hline
     Component & Parameter (unit) & Value \\      
     \hline
     ${\rm FE_{low}}$ & $kT_e$ (keV) & 0.09 (fixed) \\
     & $Z_{\rm all}$ (solar) & 0.05 (fixed)\\
     & VEM ($10^{59}~{\rm cm^{-3}}$)$^{\dagger}$ & 1.8 (fixed)\\
     ${\rm FE_{high}}$ & $kT_e$ (keV) & 0.59 (fixed) \\
     & $Z_{\rm all}$ (solar) & 0.05 (fixed)\\
     & VEM ($10^{56}~{\rm cm^{-3}}$)$^{\dagger}$ &  $6.7$ (fixed)\\      \hline
     CXB & $N_{\rm H}$~(10$^{22}$~cm$^{-2}$) & 8.44 (fixed) \\
     & Photon index & 1.4 (fixed) \\
     & Normalization$^{\ddagger}$ & 9.69 (fixed) \\  \hline
     ${\rm LP}$ & $kT_e$ (keV) & 1.33 (fixed) \\
     & $Z_{\rm Ar}$ (solar)& 1.07 (fixed)\\
     & $Z_{\rm other}$ (solar) & 0.81 (fixed)\\
     & VEM ($10^{56}~{\rm cm^{-3}}$)$^{\dagger}$ &$6.5\pm0.7$\\
      ${\rm HP}$ & $kT_e$ (keV) & $7.5^{+2.7}_{-1.7}$ \\
     & $Z_{\rm Ar}$ (solar) & 1.07 (fixed)\\
     & $Z_{\rm other}$ (solar) & 0.81  (fixed)\\
     & VEM ($10^{56}~{\rm cm^{-3}}$)$^{\dagger}$ & 0.29 $\times$ {LP VEM} (fixed)\\ 
   \hline
    ${{{\chi_\nu}^2}~(\nu)}^\S$ & & 1.42 (76) \\
     \hline
   \end{tabular}}
   \label{tab:bkg_model}
\begin{tabnote}
$^{\dagger}$Volume emission measure VEM = $\int \,n_e\,n_{\rm H}\,dV$, where $n_e$,  $n_ {\rm H}$, and $V$ are the electron density, the hydrogen density, and the emitting volume, respectively.\\
$^{\ddagger}$The unit is $\rm photons~s^{-1}~cm^{-2}~keV^{-1}~sr^{-1}$ at 1~keV.\\
$^{\S}$ The parameters ${\chi_\nu}^2$ and $\nu$ indicate a reduced chi square and a degree of freedom, respectively.
\end{tabnote}
\end{table}

\begin{table*}
 \tbl{The 4.0--9.0 keV surface brightness and intensities and of Fe lines$^{\dagger}$.}{
 \begin{tabular}{llllllll}
     \hline
     line & outside & region 1 &region  2 & region 3 & region 4 \\      
     \hline
     	Fe\emissiontype{I} K$\alpha$ & $0.33^{+0.19}_{-0.20}$ & $0.64^{+0.24}_{-0.22}$ & $0.41^{+0.18}_{-0.17}$ & $0.40\pm0.16$ & $0.23^{+0.11}_{-0.09}$\\
	Fe\emissiontype{XXV} He$\alpha$ & $0.74\pm0.22$ & $0.81^{+0.29}_{-0.24}$ &  $0.84^{+0.19}_{-0.20}$ & $0.94\pm0.18$ & $1.48\pm0.15$\\
	      \hline
   \end{tabular}}
   \label{tab:Fe}
\begin{tabnote}
$^{\dagger}$The units is 10$^{-7}~\rm photons\,s^{-1}\,cm^{-2}\,arcmin^{-2}$.
\end{tabnote}
\end{table*}

\subsection{Spectral Analysis}\label{sec:speana}
\begin{figure*}
\begin{tabular}{cccc}
\begin{minipage}[c]{0.5\hsize}
\includegraphics[width=7cm]{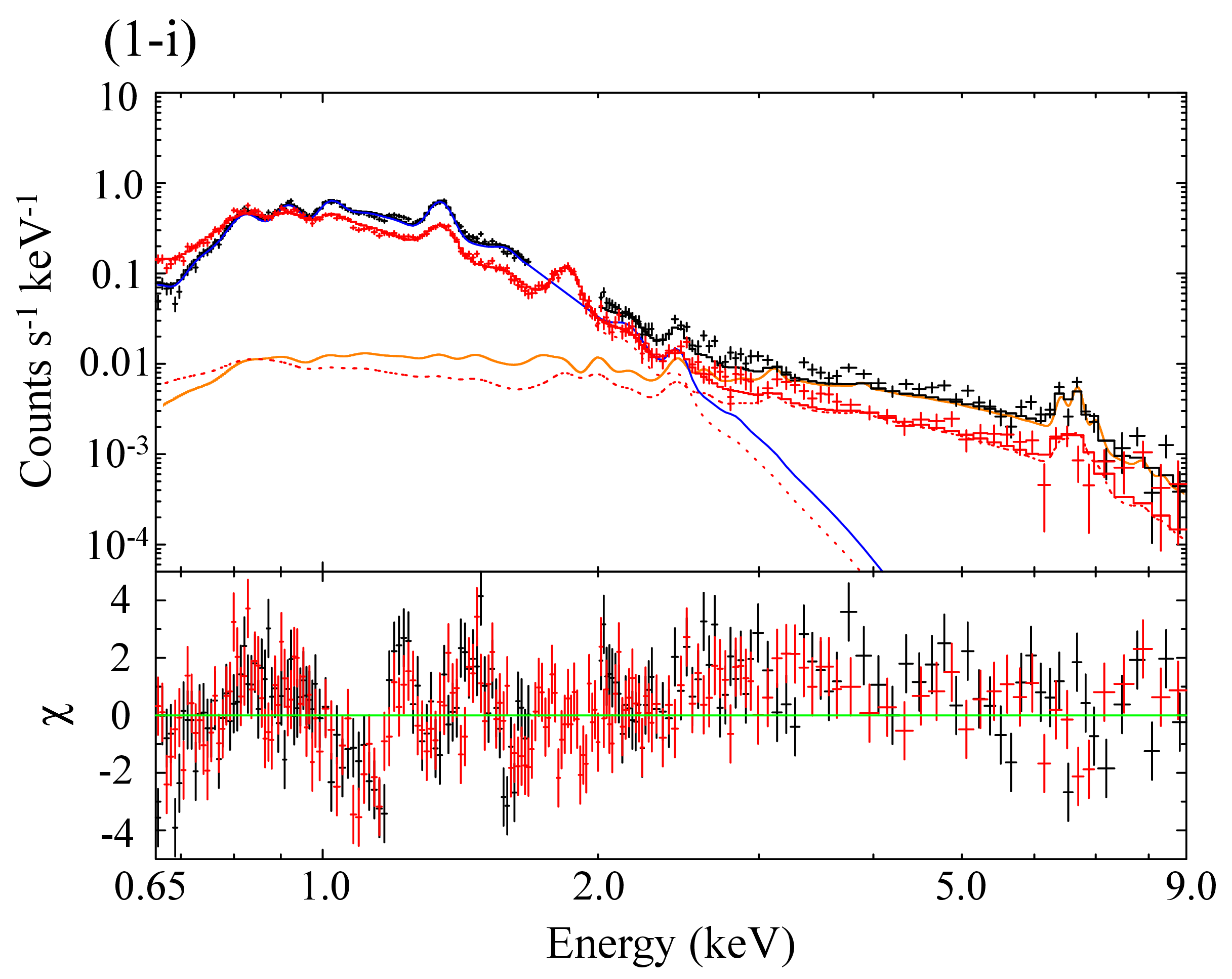} 
\end{minipage}
\begin{minipage}[c]{0.5\hsize}
\includegraphics[width=7cm]{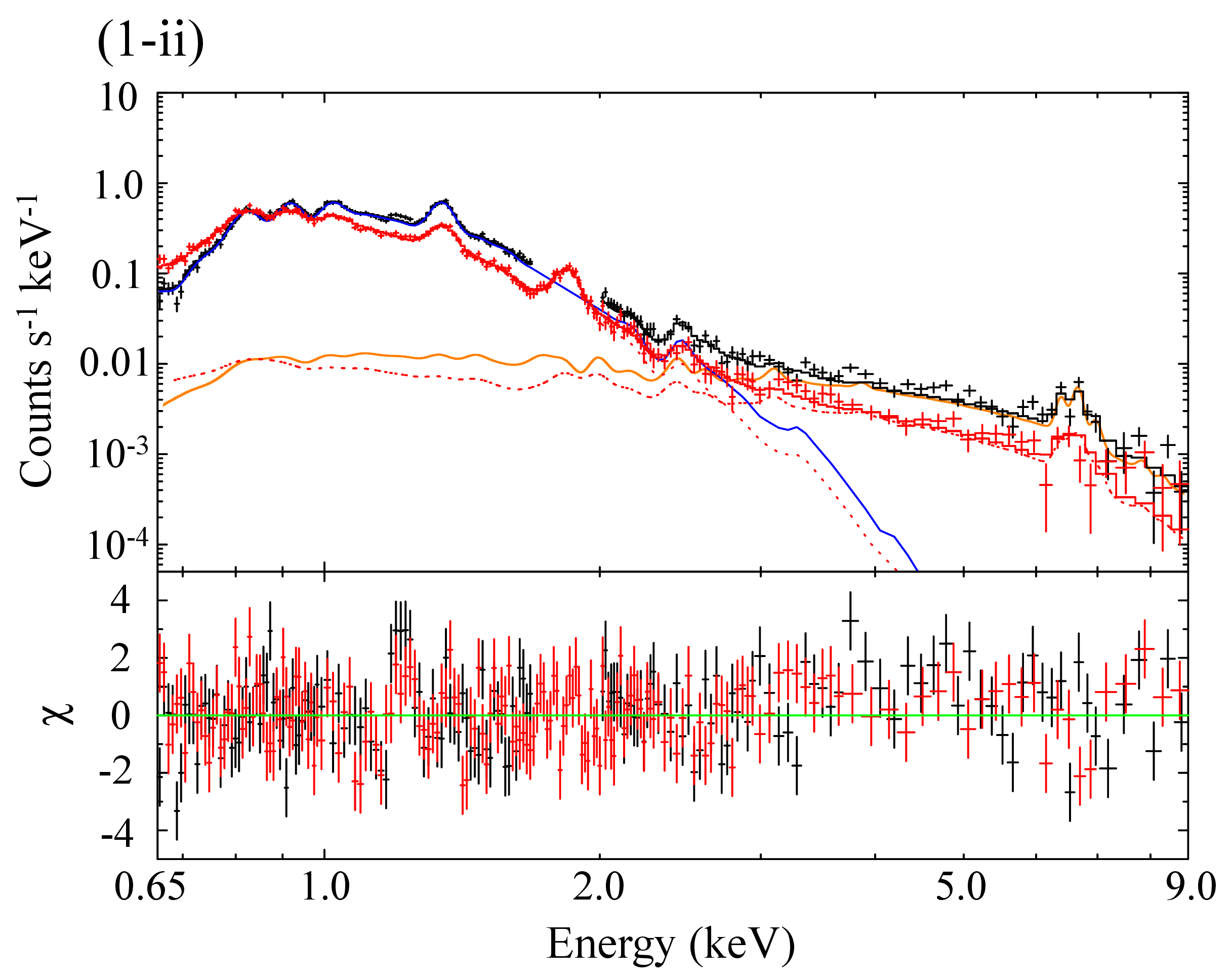} 
\end{minipage}
\\\\
\begin{minipage}[c]{0.5\hsize}
\includegraphics[width=7cm]{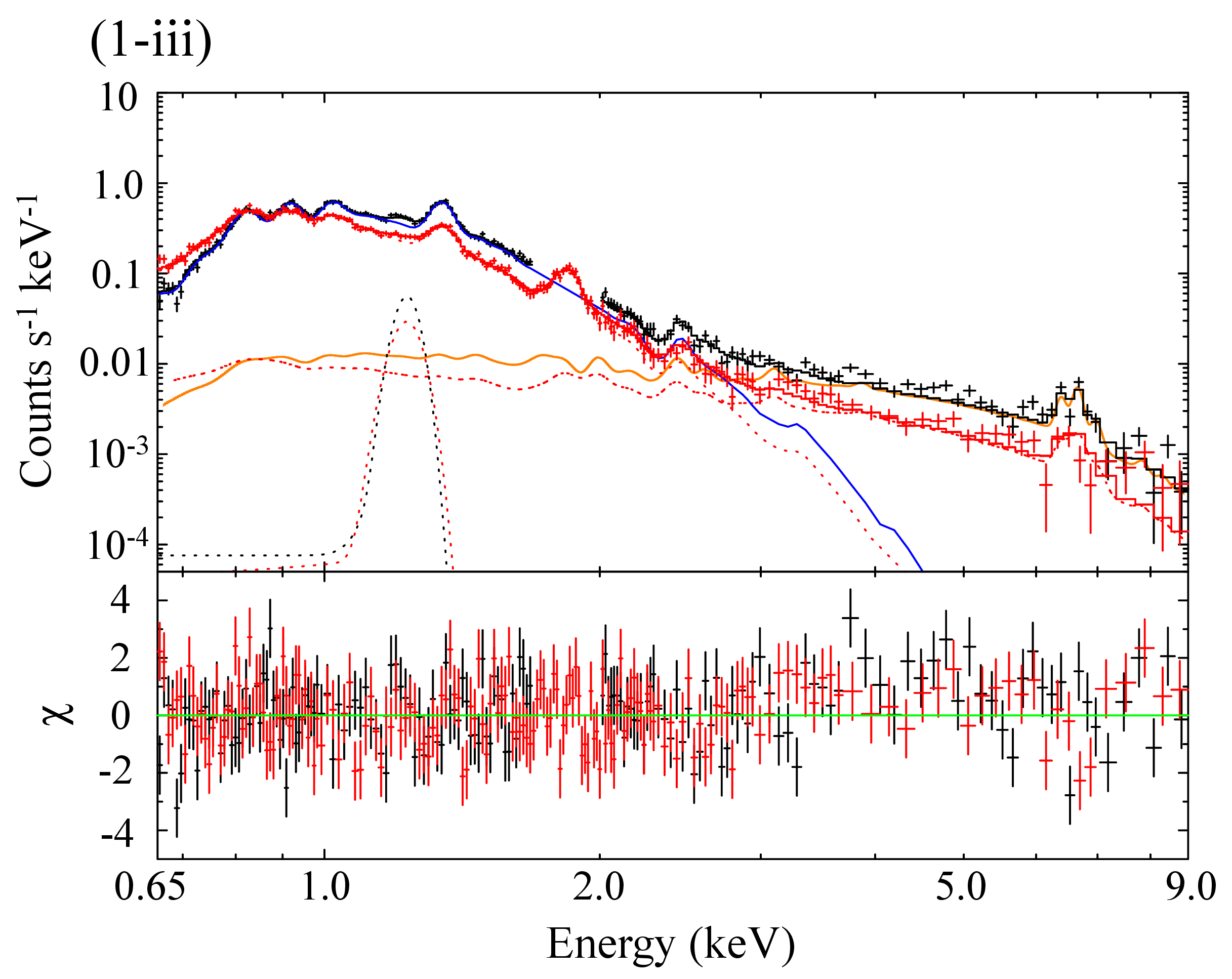} 
\end{minipage}
\begin{minipage}[c]{0.5\hsize}
\includegraphics[width=7cm]{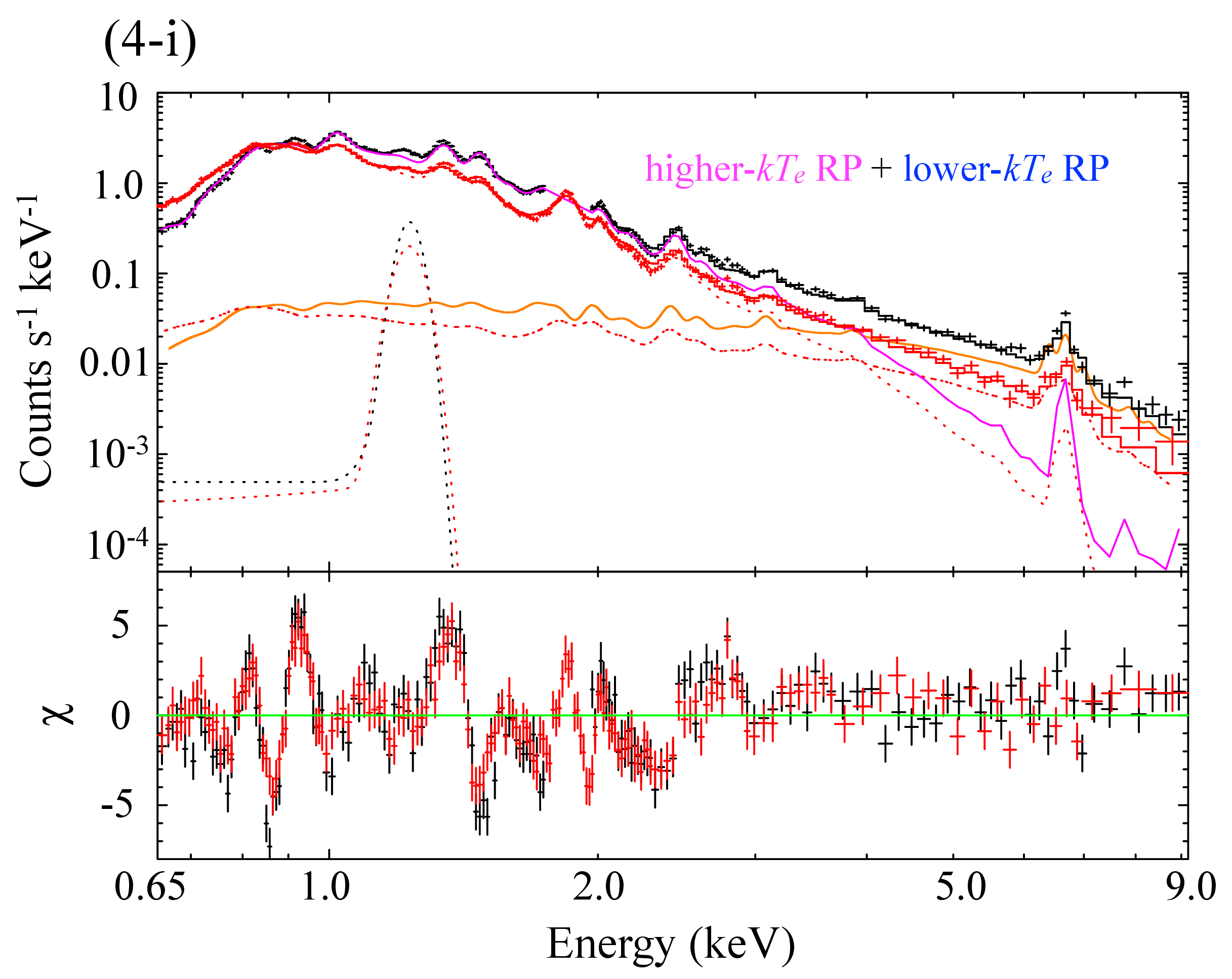}
\end{minipage}
\\\\
\begin{minipage}[c]{0.5\hsize}
\includegraphics[width=7cm]{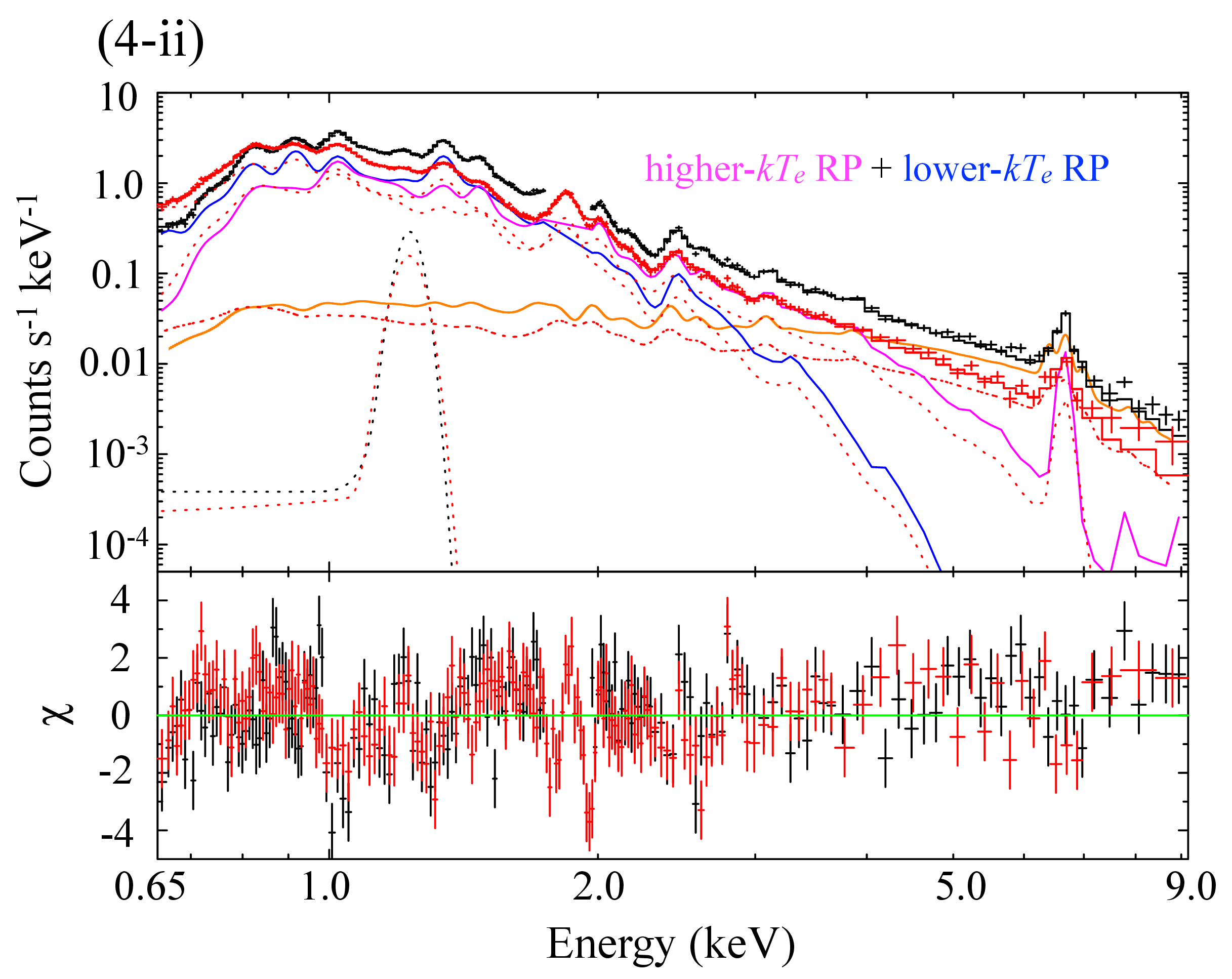} 
\end{minipage}
\begin{minipage}[c]{0.5\hsize}
\includegraphics[width=7cm]{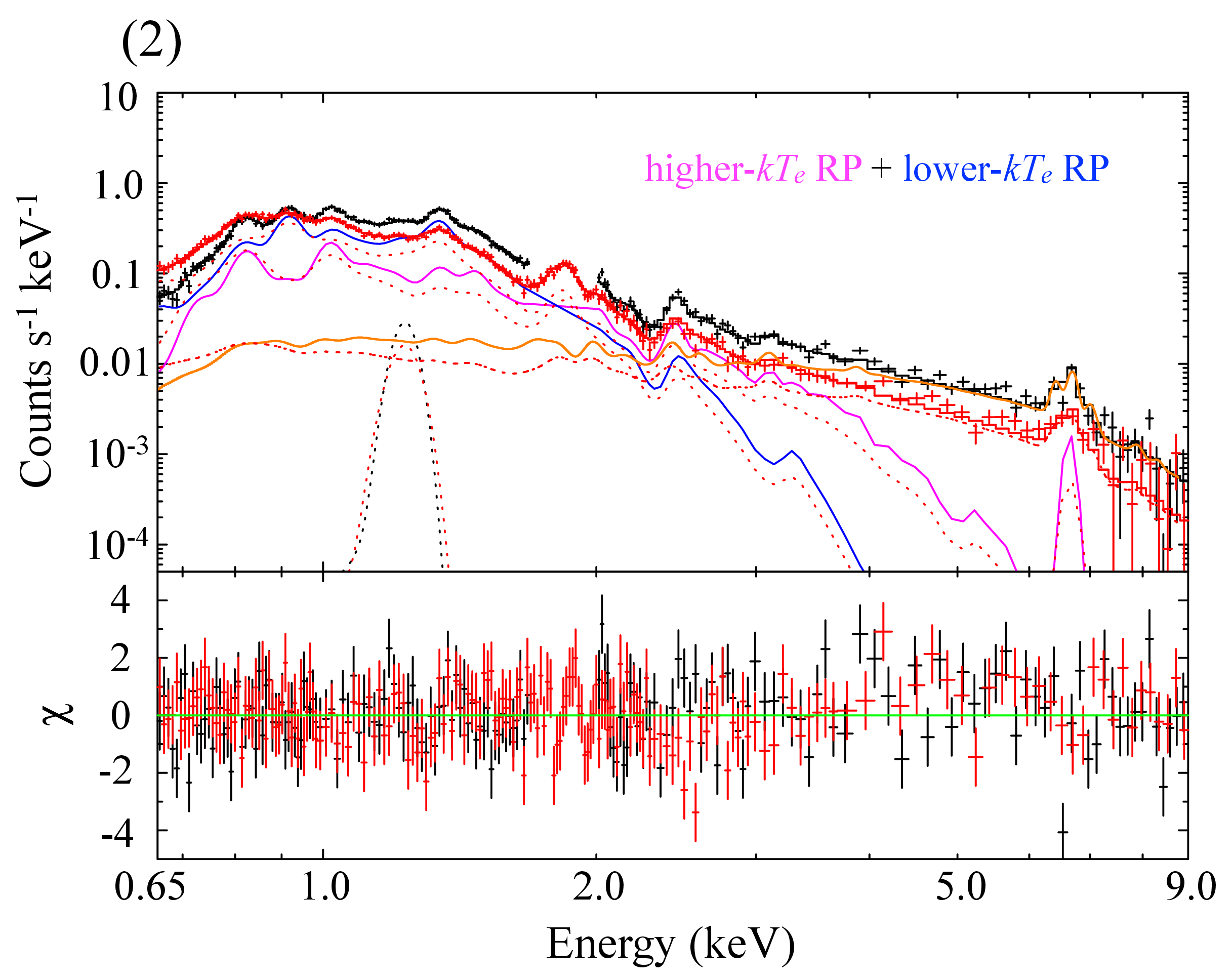} 
\end{minipage}
\\\\
\begin{minipage}[c]{0.5\hsize}
\includegraphics[width=7cm]{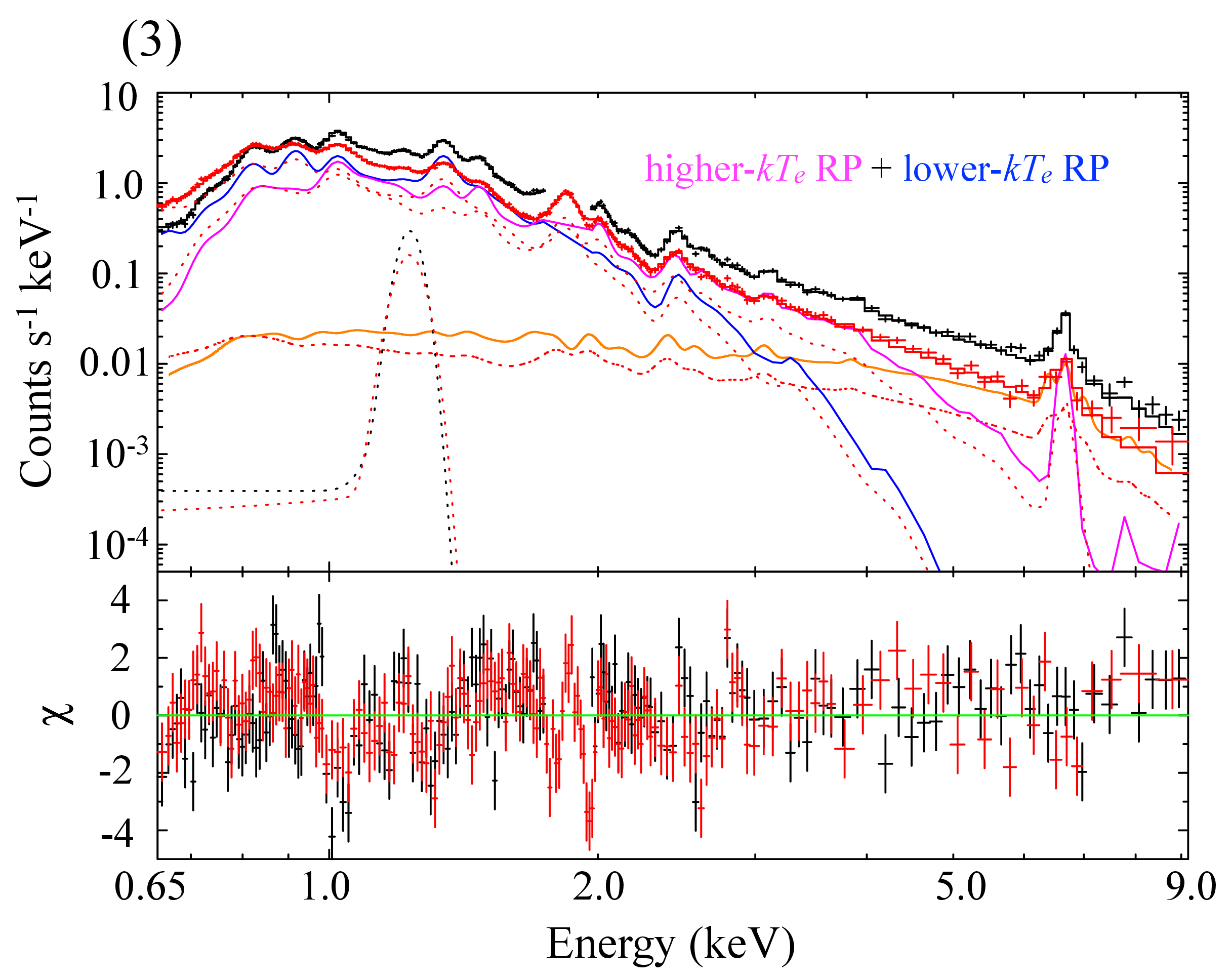} 
\end{minipage}
\end{tabular}
\vspace{+4mm}
\caption{Spectra of regions 1, 2, 3, and 4 in the energy band of 0.65--4.0~keV.
The black and red data points show the XIS0+3 (FI) and XIS1 (BI) spectra, respectively. 
The magenta, blue and orange curves represent the higher-$kT_e$ RP, the lower-$kT_e$ RP and the background for the FI spectra, respectively. 
The dotted curves are the Gaussian for the Fe-L lines at 1.23~keV.}
\label{fig:spectra_analysis}
\end{figure*}

Following the results by \citet{Zhou2014} and \citet{Nakamura2014} on the northeastern rim, 
we first tried an absorbed non-equilibrium ionization (NEI) model for the spectra of region~1.
We applied the Wisconsin absorption model \citep{Morrison1983} with solar abundances \citep{Andres1989} with the column density $N_{\rm H}$ allowed to vary
The electron temperature $kT_e$, ionization time scale $n_et$ and normalization of the NEI component were allowed to vary.
The abundances of Ne, Mg, Si, S and Fe were also set free, whereas Ar and Ca were linked to S, and Ni was linked to Fe.
The abundances of all the other elements were fixed to solar.
As explained in section~\ref{sec:lxb}, the parameters of the background model were fixed to the values presented in table~\ref{tab:bkg_model}, while the normalization of the Fe\emissiontype{I}-K lines was allowed to vary.
Note that the best-fit values of the Fe\emissiontype{I}-K lines are consistent with those displayed in figure~\ref{fig:Fe_emission} and in table~\ref{tab:Fe}.
Figure~\ref{fig:spectra_analysis}~(1-i) shows the result.
The result yields a large $n_et\geq3 \times 10^{12}~\rm{cm^{-3}\,s}$, which prefers a collisional ionization equilibrium (CIE; \cite{Masai1994}).
However, this model is statistically unacceptable (${\chi_\nu}^2$ = 2.16 with $\nu$ = 341).
We found that large residuals are still seen below 2~keV, especially at $\sim1.5$~keV around the Mg\emissiontype{XII} Ly$\alpha$ line, indicating that a higher ionization state is required for the plasma in region~1.

We therefore applied an RP model, where an initial temperature $kT_{\rm init}$ was allowed to vary in addition to the above free parameters.
We linked S to Si since these abundances are consistent within the 90\% error margin.
As shown in figure~\ref{fig:spectra_analysis}~(1-ii), the RP model considerably improved the residuals below 2~keV other than an excess 1.2~keV line (${\chi_\nu}^2$ =1.44 with $\nu$ = 339).
We consider that the excess is due to the lack of Fe-L lines in a current plasma code, as pointed out by a number of previous studies (e.g., \cite{Hughes1998}, \cite{Borkowski2006}, \cite{Yamaguchi2011}).
Adding a Gaussian at 1.23~keV, we refited the spectrum and obtained a statistically good fit as shown in figure~\ref{fig:spectra_analysis}~(1-iii) (${\chi_\nu}^2$ = 1.31 with $\nu$ = 338).
The best-fit parameters are listed in table~\ref{tab:spectra_analysis}.
Note that our result is inconsistent with the results of XMM-Newton \citep{Zhou2014, Nakamura2014}.
This is simply because the excess of the Mg\emissiontype{XII} Ly$\alpha$ emission we found is relatively weak, which cannot be correctly measured in the statistically poorer ($<30$~ks) EPIC spectra they analyzed.

We next analyzed the spectrum of region~4.
Following the result given by \citet{Sawada2012}, we also fitted the spectrum with a similar RP model as explained above.
With the RP model, we cannot reproduce the spectrum (${\chi_\nu}^2$ = 5.15 with $\nu$ = 346).
In figure~\ref{fig:spectra_analysis} (4-i), large residuals appear below 3~keV, particularly at $\sim0.9$~keV and $\sim1.0$~keV, corresponding to the Ne\emissiontype{IX} He$\alpha$ and the Ne\emissiontype{X}  Ly$\alpha$ lines, respectively.
On the other hand, the model gives a fairly good fit above 3~keV.
These results naturally indicate that required ion populations (i.e., ionization states) are different between the soft-band and hard-band spectra.
We therefore tried a two-RP model with different ionization states; neither an NEI$+$RP model nor a CIE$+$RP model can reproduce 
the spectrum particularly below 1~keV  (${\chi_\nu}^2$ = 2.03 with $\nu$ = 336 and ${\chi_\nu}^2$ = 2.04 with $\nu$ = 338, respectively).
We linked the abundances and $kT_{\rm init}$ of the two components because these parameters are within 90\% error margin when we thawed them.
As shown in figure~\ref{fig:spectra_analysis}~(4-ii) and summarized in table~\ref{tab:spectra_analysis}, the two-RP model explained well the spectrum of region~4 (${\chi_\nu}^2$ = 1.71 with $\nu$ = 343).

Since regions~2 and 3 are located between region~1 and region~4, we expect that their spectra should be described by a similar model with those for regions~1 or 4.
We thus tried two different models: single RP and two-RP.
The single-RP model results in large residuals in the soft energy band ($<1$~keV) whereas the two-RP model gives a statistically acceptable fit, as displayed in figures~\ref{fig:spectra_analysis}~(2) and (3).
The detailed parameters are summarized in table~\ref{tab:spectra_analysis}.

\begin{table*}[h!]
 \tbl{Best-fit model parameters of SNR plasmas.}{
 \begin{tabular}{llllllll}
     \hline
     Model function & Parameter (unit) & (region 1) & (region 2) & (region 3) & (region 4) \\      
     \hline
     Absorption & $N_{\rm H}$ (10$^{22}$ cm$^{-2}$) & $0.88^{+0.02}_{-0.03}$ & $0.89^{+0.03}_{-0.02}$  & $0.83^{+0.03}_{-0.02}$ & $0.75\pm0.01$\\
     VVRNEI1 & $kT_e$ (keV) & $0.247^{+0.05}_{-0.07}$ & $0.18\pm0.01$ & $0.179^{+0.011}_{-0.004}$ & $0.216^{+0.001}_{-0.005}$\\
     & $kT_{\rm init}$ (keV) & $\geq2.4$ & $3.7^{+1.1}_{-0.9}$ & $3.5^{+1.0}_{-0.8}$ & $3.8^{+0.4}_{-0.6}$ \\
     & $Z_{\rm Ne}$ (solar) & $1.3\pm0.1$ & $1.7^{+0.2}_{-0.3}$ & $2.0^{+0.2}_{-0.1}$ & $1.73^{+0.07}_{-0.06}$\\
     & $Z_{\rm Mg}$ (solar) & $1.14^{+0.09}_{-0.08}$ & $1.3\pm0.2$ & $1.5\pm0.1$ & $1.57^{+0.05}_{-0.04}$ \\
     & $Z_{\rm Si}$ (solar)  & $1.1\pm0.1$ & $1.3\pm0.2$ & $1.3\pm0.2$ & $1.4\pm0.1$ \\
     & $Z_{\rm S}$ = $Z_{\rm Ar}$ = $Z_{\rm Ca}$ (solar) & = $Z_{\rm Si}$ & $1.7^{+0.3}_{-0.2}$ & $1.6\pm0.3$ & $1.38\pm0.07$ \\
     & $Z_{\rm Fe}$ = $Z_{\rm Ni}$ (solar) & $1.0\pm0.1$ & $2.4\pm0.7$ & $2.9\pm0.3$ & $1.17^{+0.03}_{-0.05}$ \\
     & $n_et$ (10$^{11}$ ${\rm cm^{-3}\,s}$) & $12.5^{+0.01}_{-0.02}$ & $11.8^{+1.2}_{-0.7}$ & $10.2^{+0.8}_{-0.4}$ & $10.1\pm0.2$ \\
     & VEM ($10^{57}~{\rm cm^{-3}}$)$^{\dagger}$ & $5.3^{+0.4}_{-0.6}$ & $8^{+1}_{-2}$ & $11^{+0.3}_{-0.1}$ & $16^{+1}_{-1}$  \\
     VVRNEI2 & $kT_e$ (keV) & - & $0.28^{+0.03}_{-0.02}$ & $0.43^{+0.21}_{-0.07}$ & $0.52^{+0.01}_{-0.02}$ \\
     & $n_et$ (10$^{11}~{\rm cm^{-3}\,s}$) & - & $4.4\pm0.8$ & $4.0^{+0.5}_{-0.6}$ & $3.7^{+0.2}_{-0.1}$ \\
     & VEM ($10^{57}~{\rm cm^{-3}}$)$^{\dagger}$ & - & $0.7^{+0.3}_{-0.2}$ & $0.4^{+0.1}_{-0.2}$ & $2.2\pm0.2$ \\
     Gaussian & Centroid (keV) & 1.23 (fixed) & 1.23 (fixed) &1.23 (fixed) & 1.23 (fixed) \\
     & Normalization$^{\ddagger}$ & $1.6\pm{0.3}$ & $0.6\pm0.3$ & $0.7\pm0.2$ & $1.6\pm{0.2}$ \\
     \hline
     ${\chi_\nu}^2$ ($\nu$)  & &  1.31 (338) & 1.15 (347) & 1.58 (343) & 1.71 (343) \\
     \hline
   \end{tabular}}
   \label{tab:spectra_analysis}
\begin{tabnote}
$^{\dagger}$Volume emission measure VEM=$\int\,n_e\,n_{\rm H}\,dV$, where $n_e$,  $n_{\rm H}$, and $V$ are the electron density, the hydrogen density, and the emitting volume, respectively.\\
$^{\ddagger}$The unit is 10$^{-6}~\rm photons\,s^{-1}\,cm^{-5}\,arcmin^{-2}$.
\end{tabnote}
\end{table*}

\section{Discussion}
\begin{figure}
\begin{center}
\includegraphics[width=8cm]{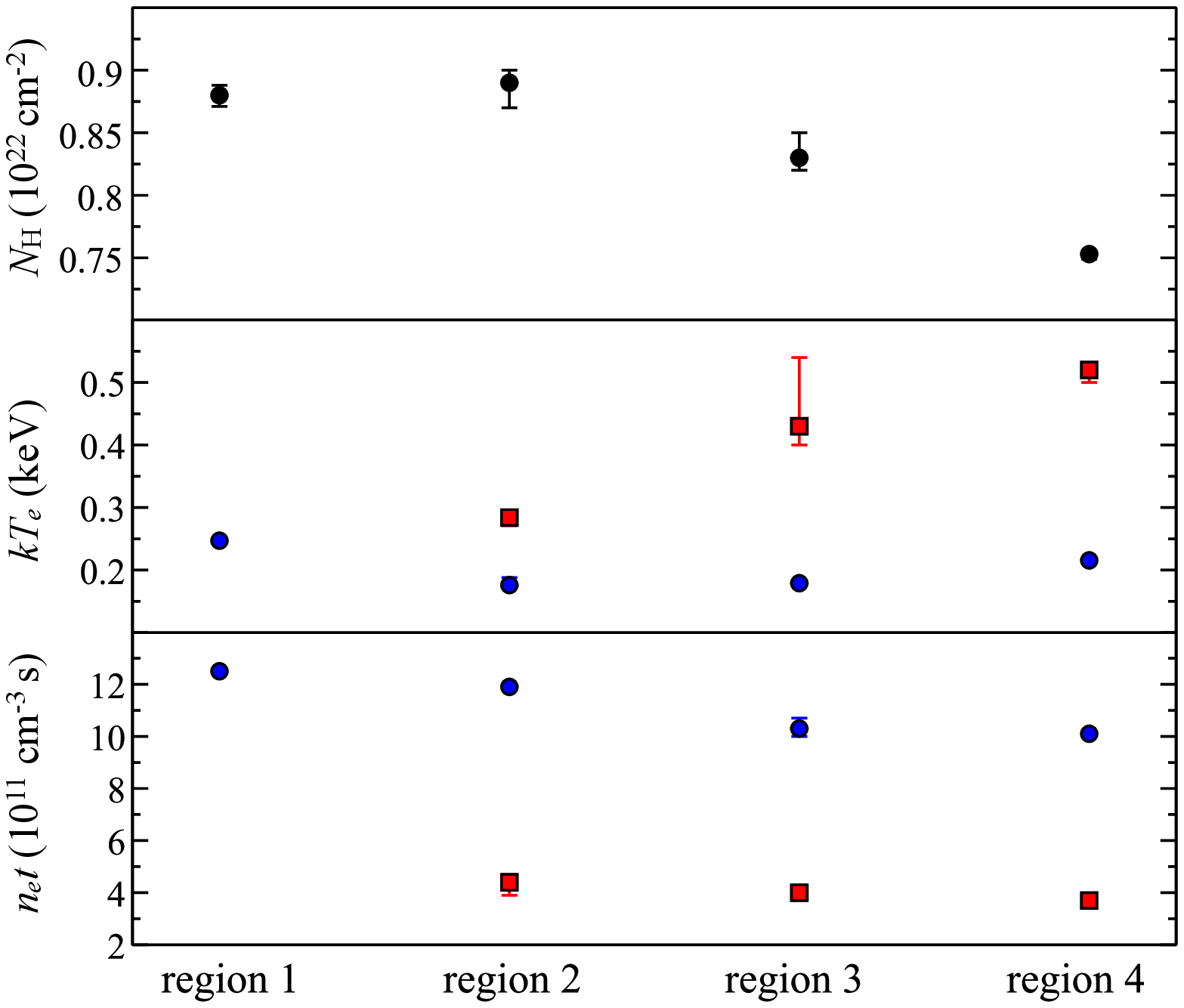} 
\end{center}
\caption{$N_{\rm H}$ (upper), $kT_e$ (middle) and $n_et$ (lower) of the two RP components obtained for the four regions.
The red and blue points represent the parameters for the higher-$kT_e$ RPs and those for the lower-$kT_e$ RPs, respectively.}
\label{fig:kTe_net}
\end{figure}

As best fit in the previous section, the X-ray spectra are explained by a single RP (region~1: the northeastern rim) or two-component RP (regions~2, 3 and 4: the inside) models.
Based on the best-fit results, we plot the absorption column density $N_{\rm H}$, electron temperature $kT_e$, and ionization timescale $n_et$ of the RP components in figure~\ref{fig:kTe_net}.

\subsection{Gas Environment of W28}
As indicated in figure~\ref{fig:kTe_net}, the absorption column density $N_{\rm H}$ is significantly higher in regions~1 and 2 than those in regions~3 and 4.
The values are consistent with the previous X-ray studies for each region \citep{Pannuti2017, Nakamura2014, Zhou2014}.
The spatial variation of $N_{\rm H}$ roughly correlates with the $\mathrm{^{12}CO~({\it J}=2-1)}$ emission and the 1720~MHz OH maser spots (figure~\ref{fig:W28_Xray_molecularclouds}).
A position-velocity diagram of $\mathrm{CO~({\it J}=3-2)}$ given by \citet{Arikawa1999} suggests that the molecular cloud is associated with W28 and is situated in front of the remnant.
According to \citet{Aharonian2008}, the ${\rm^{12}CO }~({\it J}=1-0)$ observation with NANTEN2 indicates that a part of the cloud is in front of the remnant in the northeastern region.
Our X-ray analysis confirms their results: the northeastern part (regions~1 and 2) of W28 is largely obscured by the molecular cloud interacting with the remnant.

\subsection{Origin of Neutral Fe-K Line}\label{sec:Fe_line}
As shown in figure~\ref{fig:Fe_emission}, we detected a line excess at 6370$^{+30}_{-15}$~eV in region~1, where W28 is interacting with the molecular cloud.
Although one possible origin of this line is thermal emission from the remnant, it is less plausible because W28 is one of the middle-aged and core-collapse SNRs, from which K$\alpha$ lines of nearly-neutral Fe are generally not expected \citep{Yamaguchi2014}.
An alternative possibility is that accelerated cosmic-ray particles bombard the northeastern cloud and emit the enhanced Fe\emissiontype{I} K$\alpha$ line emissions. 
A similar interpretation was proposed by \citet{Sato2014} and \citet{Sato2016} in the cases of two other other mixed-morphology SNRs, 3C~391 and Kes~79, respectively.
Since GeV/TeV $\gamma$-ray observations of W28 provide evidence for the presence of particle acceleration \citep{Abdo2010,Aharonian2008}, this scenario would be more likely.
Further evidence is provided by \citet{Nobukawa2018}, who reported that enhanced Fe\emissiontype{I} K$\alpha$ line emissions are detected from four Galactic SNRs (Kes~67, Kes~69, Kes~78 and W44) and also from the central region of W28.

\subsection{Recombining Plasma}
As presented in section~\ref{sec:speana}, our results clearly suggest that two RP plasmas with different temperatures are required for the inner regions (2, 3, and 4).
From figure~\ref{fig:kTe_net}, the ``cold'' component has an electron temperature $kT_e\sim0.2~{\rm keV}$ whereas the ``hot'' component has a much higher temperature $0.3$--$0.5~{\rm keV}$.
The temperature of the hot component in region~2 is close to that of the cold component and is the lowest among the three inner regions.
On the other hand, the spectrum of region~1 is adequately represented by a single RP component with $kT_e\sim0.25~{\rm keV}$.
It is notable that the temperature in region~1 is close to the average temperature in region~2.
Taking account of the gas environment discussed in the previous section, we presume that these results hint at the physical structure of the plasma in W28: the temperature in the northeast has decreased due to the dense surrounding environment which gradually cools the remnant toward the center.

A similar picture was proposed by \citet{Matsumura2017b}, who indicated that IC~443 is cooled in the southeastern region where the blast waves are in contact with molecular clouds. 
We point out that both W28 and IC~443 are the MM SNRs largely dominated by RPs.
Since all the RP SNRs discovered so far are classified as MM SNRs (see table~4 in \cite{Uchida2015}) that show strong evidence for SNR-cloud interaction \citep{Rho1998}, it is reasonable to assume thermal conduction as a possible scenario to create the RP. 
As discussed by \citet{Kawasaki2002}, the RP can arise from thermal conduction between the hot ejecta and dense surrounding environment 
when the cooling timescale is shorter than the recombination timescale.
While \citet{Sato2014} and \citet{Washino2016} focused on thermal conduction between the downstream shocked material and the upstream neutral gas, it is generally difficult to calculate an accurate timescale since a saturated conduction may not be negligible near  the contact surface of the shock front \citep{Dalton1993}.  
On the other hand, \citet{White1991} proposed a cloud evaporation model to account for the center-filled morphology of MM SNRs.
\citet{Zhou2011} also presumed that the northeastern rim is a product of cloud evaporation.
If this is the case, the evaporation scenario may be a potential origin of the overionization.
However, current simulations are too simple to predict the spectrum expected in this scenario; to discuss the true origin of RPs, the inclusion of more physical processes (e.g., non-equilibrium ionization) is required in future works \citep{Slavin2017}.

\citet{Sawada2012} favored another plausible origin of the RP, namely rarefaction \citep{Itoh1989,Shimizu2012}.
According to several calculations \citep{Itoh1989,Shimizu2012}, the blast wave should have broken out of the circumstellar material 
less than several 100~yr after the explosion \citep{Moriya2012}.
The timescale is much shorter than the estimated age of W28, 33--42~kyr \citep{Velazquez2002,Kaspi1993,Li2010}.
From the best-fit value of $n_e t$ presented in table~\ref{tab:spectra_analysis}, we estimated the elapsed time since the rarefaction to be $11\times(n_e\,/\,1~{\rm cm^{-3}})^{-1}$~kyr and $32\times(n_e\,/\,1~{\rm cm^{-3}})^{-1}$~kyr for the high-$kT_e$ RP and the low-$kT_e$ RP in region 4, respectively, 
We note that the result is roughly consistent with the estimation given by \citet{Sawada2012}, $\sim10\times(n_e\,/\,1~{\rm cm^{-3}})^{-1}$~kyr.
These timescales are on the same order of the age of W28, indicating that the rarefaction scenario cannot be ruled out.
In this scenario, however, the spatial variations of temperature and ionization state are difficult to explain, as previously noted by \citet{Sawada2012}.

\section{Conclusion}
We analyzed the Suzaku XIS data of the central and northeastern regions of W28.
In the spatially resolved analysis, we found that each spectrum is best explained by an RP model.
The X-ray spectra in the inner regions are well reproduced by a combination of two RPs; a cold low-ionized ($kT_e\sim0.2~{\rm keV}$, $n_et\sim10^{12}~{\rm cm^{-3}\,s}$) and a hot highly ionized ($kT_e\geq0.3~{\rm keV}$, $n_et\sim4\times10^{11}~{\rm cm^{-3}\,s}$) components.
On the other hand, the spectrum of the northeastern rim is represented by a single RP with $kT_e\sim0.25~{\rm keV}$.
The spatial variations of these parameters suggest that a thermal conduction (e.g., cloud evaporation scenario) is the origin of the overionization, although we cannot rule out the rarefaction scenario as an alternative scenario.
We also found the Fe\emissiontype{I} K$\alpha$ line only in the northeastern rim region, suggesting possible interactions between cosmic-ray particles and the molecular cloud.

\begin{ack}
We are grateful to Dr. Makoto Sawada for helpful advice.
We thank Dr. Satoshi Yoshiike for providing us with the NANTEN2 data.
We deeply appreciate all the Suzaku team members. 
This work is supported by JSPS/MEXT Scientific Research Grant Numbers JP15J01842 (H.M.), JP25109004 (T.T. and T.G.T.),  JP26800102 (H.U.), JP15H02090 (T.G.T.), and JP26610047 (T.G.T.).
\end{ack}

\end{document}